\documentclass[10pt,letterpaper,english,conference]{IEEEtran}
\usepackage[T1]{fontenc}
\usepackage[latin9]{inputenc}
\usepackage{amsmath}
\usepackage{graphicx}
\usepackage{esint}

\makeatletter

\pdfpageheight\paperheight
\pdfpagewidth\paperwidth

\providecommand{\tabularnewline}{\\}

\usepackage{flushend}
\usepackage{tikz}
\usetikzlibrary{arrows,shadows,positioning,backgrounds}
\usepackage{algorithm}
\usepackage{algpseudocode}
\usepackage{cite}

\usepackage{balance}

\makeatother

\usepackage{hyperref}
\usepackage{lipsum}
\newcommand\copyrighttext{%
	\footnotesize \textcopyright 2020 IEEE. Personal use of this material is permitted.
	Permission from IEEE must be obtained for all other uses, in any current or future
	media, including reprinting/republishing this material for advertising or promotional
	purposes, creating new collective works, for resale or redistribution to servers or
	lists, or reuse of any copyrighted component of this work in other works.
	DOI: \href{https://doi.org/10.1109/CCNC49032.2021.9369496}{10.1109/CCNC49032.2021.9369496}}
\newcommand\copyrightnotice{%
	\begin{tikzpicture}[remember picture,overlay]
		\node[anchor=south,yshift=10pt] at (current page.south) {\fbox{\parbox{\dimexpr\textwidth-\fboxsep-\fboxrule\relax}{\copyrighttext}}};
	\end{tikzpicture}%
}

\usepackage{babel}
\begin{document}
\title{Reducing FEC-Complexity in Cross-Layer Predictable Data Communication}
\author{\IEEEauthorblockN{Pablo Gil Pereira and Thorsten Herfet\IEEEmembership{IEEE Senior Member}}\\
 \IEEEauthorblockA{Telecommunications Lab,\\
 Saarland Informatics Campus, D-66123 Saarbrücken, Germany\\
 Email: \{gilpereira, herfet\}@cs.uni-saarland.de}}
\maketitle
\copyrightnotice
\begin{abstract}
The PRRT protocol enables applications with strict performance requirements
such as Cyber-Physical Systems, as it provides predictably low, end-to-end
delay via cross-layer pacing and timely error correction via Hybrid
ARQ (HARQ). However, the implemented HARQ uses computationally complex
Maximum Distance Separable (MDS) codes to generate redundancy. In
this paper we propose code partitioning for the complexity reduction
of MDS codes, thereby enabling their deployment on constrained embedded
devices.
\end{abstract}


\thispagestyle{empty}

{\let\thefootnote\relax\footnotetext{The work is supported by the German Research Foundation~(DFG) within SPP 1914 ``Cyber-Physical Networking'' under grant 315036956}}

\section{Introduction\label{sec:Introduction}}

Cyber-Physical Systems (CPS) integrate physical and digital processes
via control loops, often using communication networks in distributed
deployments. In order to guarantee a successful operation, CPS require
predictable reliability and delay, which are difficult to achieve
due to the inherent design of computation and communication systems~\cite{lee2008cyber}.
The Predictably Reliable Real-time Transport (PRRT)~\cite{Gorius:PRRT}
protocol can optimize its configuration to meet the delay constraints
of a system with cross-layer pacing~\cite{schmidt2019cross}, which
keeps the system and network buffers empty to minimize the end-to-end
delay. Timely reliability is achieved with a Hybrid ARQ (HARQ) scheme
that extends the error correction in lower layers to cope with losses
not bound to the link within a target delay.

Since the transport layer works at the packet level, the block lengths
are usually several orders of magnitude shorter than those at the
physical layer, especially when operating under delay constraints.
Therefore, Minimum Distance Separable (MDS) codes have been used~\cite{rizzo1997effective,lacan2004systematic,soro2010fnt}
because they are more suitable for short codes than LDPC or polar
codes as they do not require excess packets. However, they have higher
complexity, which may result in large en- and decoding delays when
deployed on devices with limited computation capabilities. This paper
analyses the impact of the complexity of MDS codes on the design of
transport protocols providing predictable reliability and delay, using
PRRT as a proof of concept. Additionally, we propose code partitioning
to efficiently reduce the complexity of MDS codes by 50\% with only
a slight increase in the redundancy information.

\section{Code Partitioning\label{sec:Code-Partitioning}}

While the Automatic Repeat reQuest (ARQ) delay is mainly dominated
by the round-trip time, the inter-packet time dominates the Forward
Error Coding (FEC) delay. As a result, the information-theoretical
optimum under delay constraints will be a combination of both, which
HARQ implements~\cite{Gorius:PRRT}. Although FEC encoding and decoding
delays are usually negligible when the protocol runs on powerful machines,
this does not hold true when it runs on more constrained devices.
Both the encoding and decoding of MDS codes entail a matrix-vector
multiplication, whereas the decoding process requires an additional
matrix inversion. If systematic codes are used, MDS codes achieve
encoding complexity $\mathcal{O}(kp)$ and decoding complexity $\mathcal{O}(kp^{2})$~\cite{rizzo1997effective},
with $k$ the block length and $p$ the number of parity packets.

The algorithm to find the optimal $p$ and $k$ is outside of the
scope of this paper and the reader is referred to~\cite{Gorius:PRRT}
for more details about it. However, it is relevant to know that this
algorithm finds the largest $k$ that meets the target delay ($D_{T}$)
and packet loss rate ($PLR_{T}$) constraints because larger codes
result in lower RI than shorter ones. This property is depicted in
Table~\ref{tab:Optimum-MDS-config}, where $n=k+p$ has been calculated
such that $PLR_{FEC}\leq PLR_{T}$. $PLR_{FEC}$ is the FEC packet
loss rate as given by Eq.~\ref{eq:fec_plr}, for which we have assumed
a Binary Erasure Channel (BEC) with erasure probability $p_{e}$ and
have defined the random variable $I_{k}^{n}$, which represents the
number of packet losses in a code $C(n,k)$. Although the PLR analysis
is also available for channels with memory (i.e., Gilbert-Elliott
model), this does not change the basic findings and hence it is omitted
for clarity in this paper. Since $k$ has a major impact on the code
complexity, we propose code partitioning to split the block into two
independent blocks instead of using the largest $k$. As a result,
the used $k$ and $p$ are halved but the encoding and decoding operations
run twice for the original block length, which nevertheless reduces
the complexity to $\mathcal{O}(\frac{kp}{2})$ for encoding and $\mathcal{O}(\frac{kp^{2}}{4})$
for decoding. The packet loss rate of partitioned codes ($PLR_{FEC}^{part}$)
is given in Eq.~\ref{eq:part_plr}, where $J_{k_{1},k_{2}}$ is a
random variable representing the packet losses in a partitioned code
and $k_{max}=\left\lceil \frac{k}{2}\right\rceil $.

\begin{table}
\begin{centering}
\begin{tabular}{|c|c|c|c|c|c|c|}
\hline 
$p_{e}$ & 0.01 & 0.03 & 0.05 & 0.07 & 0.09 & 0.1\tabularnewline
\hline 
$n$ ($k=40$) & 44 & 48 & 50 & 53 & 55 & 56\tabularnewline
\hline 
$n$ ($k=80$) & 86 & 91 & 95 & 98 & 102 & 104\tabularnewline
\hline 
\end{tabular}
\par\end{centering}
~
\centering{}\caption{Optimum MDS code configuration to meet $PLR_{T}=10^{-5}$ for different
block lengths\label{tab:Optimum-MDS-config}}
\end{table}

\begin{equation}
PLR_{FEC}=\frac{1}{k}E\left[I_{k}^{n}\right]=\frac{1}{k}\sum_{i=1}^{k}i\cdot P(I_{k}^{n}=i)\label{eq:fec_plr}
\end{equation}

\begin{equation}
P(I_{k}^{n}=i)=\begin{cases}
\sum_{e=0}^{n-k}\binom{n}{e}p_{e}^{e}(1-p_{e})^{n-e} & i=0\\
\sum_{e=max(n-k+1,i)}^{n-k+i}p\binom{n}{e}\frac{\binom{k}{i}\binom{n-k}{e-i}}{\binom{n}{e}} & 1\leq i\leq k
\end{cases}\label{eq:prob_i_losses}
\end{equation}
\begin{equation}
p\binom{n}{e}=\binom{n}{e}p_{e}^{e}(1-p_{e})^{n-e}
\end{equation}

\begin{equation}
PLR_{FEC}^{part}=\frac{1}{k}E\left[J_{k_{1},k_{2}}\right]=\frac{1}{k}\sum_{j=1}^{k}j\cdot P(J_{k_{1},k_{2}}=j)\label{eq:part_plr}
\end{equation}

\begin{equation}
P(J_{k_{1},k_{2}}=j)=\sum_{e=max(0,j-k_{max})}^{min(k_{max},j)}P(I_{k_{1}}^{n_{1}}=e)\cdot P(I_{k_{2}}^{n_{2}}=j-e)\label{eq:part_prob_j}
\end{equation}

The FEC delay is impacted in two ways: i) the decrease in encoding
and decoding delays and ii) the increase in $p$ when $PLR_{FEC}^{part}>PLR_{T}\geq PLR_{FEC}$.

\section{Evaluation\label{sec:Evaluation}}

\begin{figure}
\centering{}\includegraphics[scale=0.35]{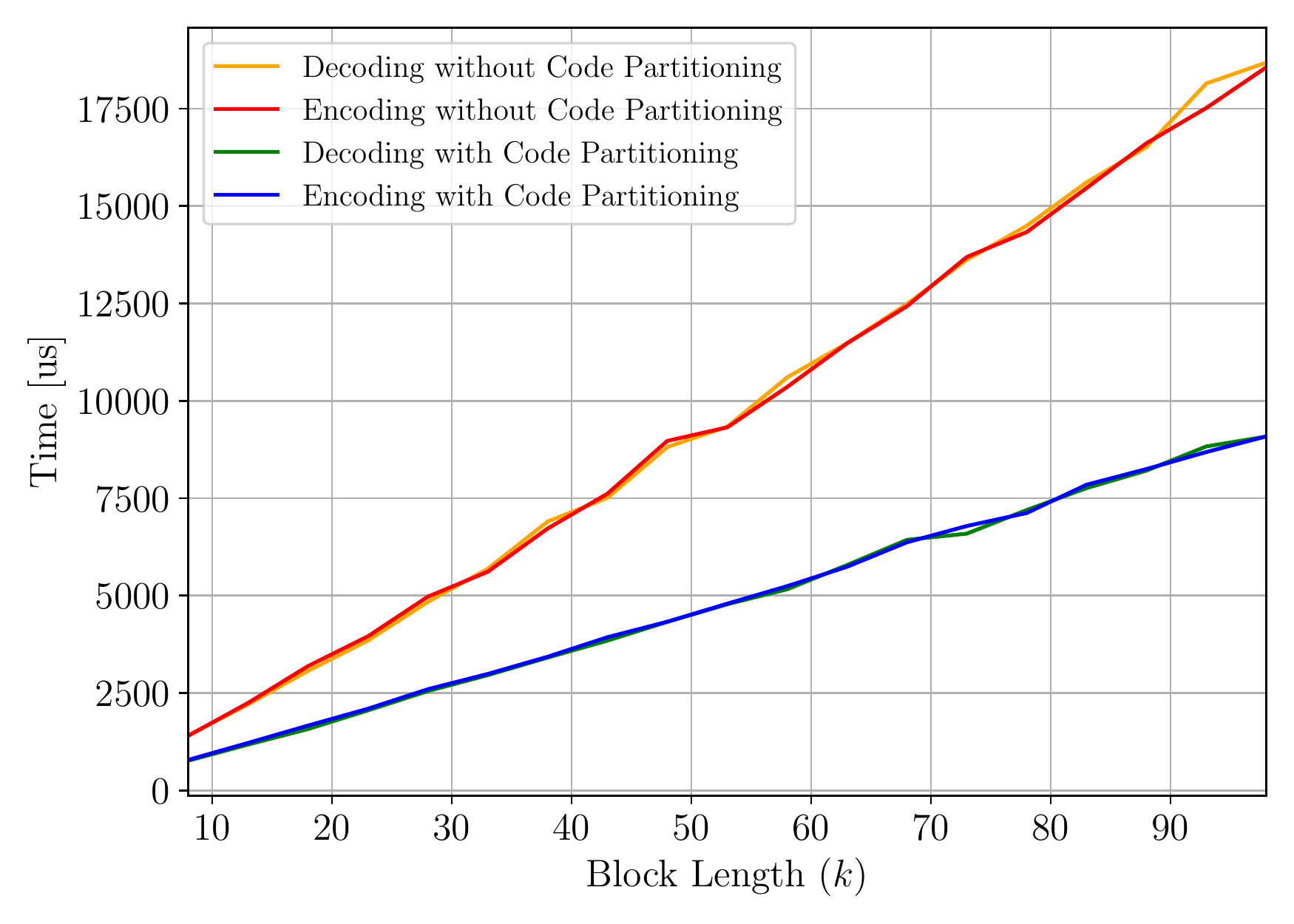}\caption{PRRT's MDS code encoding and decoding times with and without code
partitioning for $m=8$ redundancy packets.\label{fig:pi_results}}
\end{figure}

PRRT's HARQ implementation using code partitioning has been executed
on a Raspberry Pi Zero W running Raspbian Buster with Linux kernel
4.19, which has been configured to run as few processes as possible
to reduce the interference with the experiment. Figure~\ref{fig:pi_results}
shows the measurements of the encoding and decoding delays. As predicted
in Section~\ref{sec:Code-Partitioning}, the encoding delay is halved
when code partitioning is employed. A surprising result is that the
encoding and decoding delays are the same, even though the decoding
needs to perform a matrix inversion. The matrix inversion code has
been isolated, resulting in 0.25\,ms delay for $k=100$, which shows
that the decoding delay is dominated by the matrix-vector multiplication
of 1500\,B packets and the inversion is negligible.

\begin{figure}
\centering{}\includegraphics[scale=0.38]{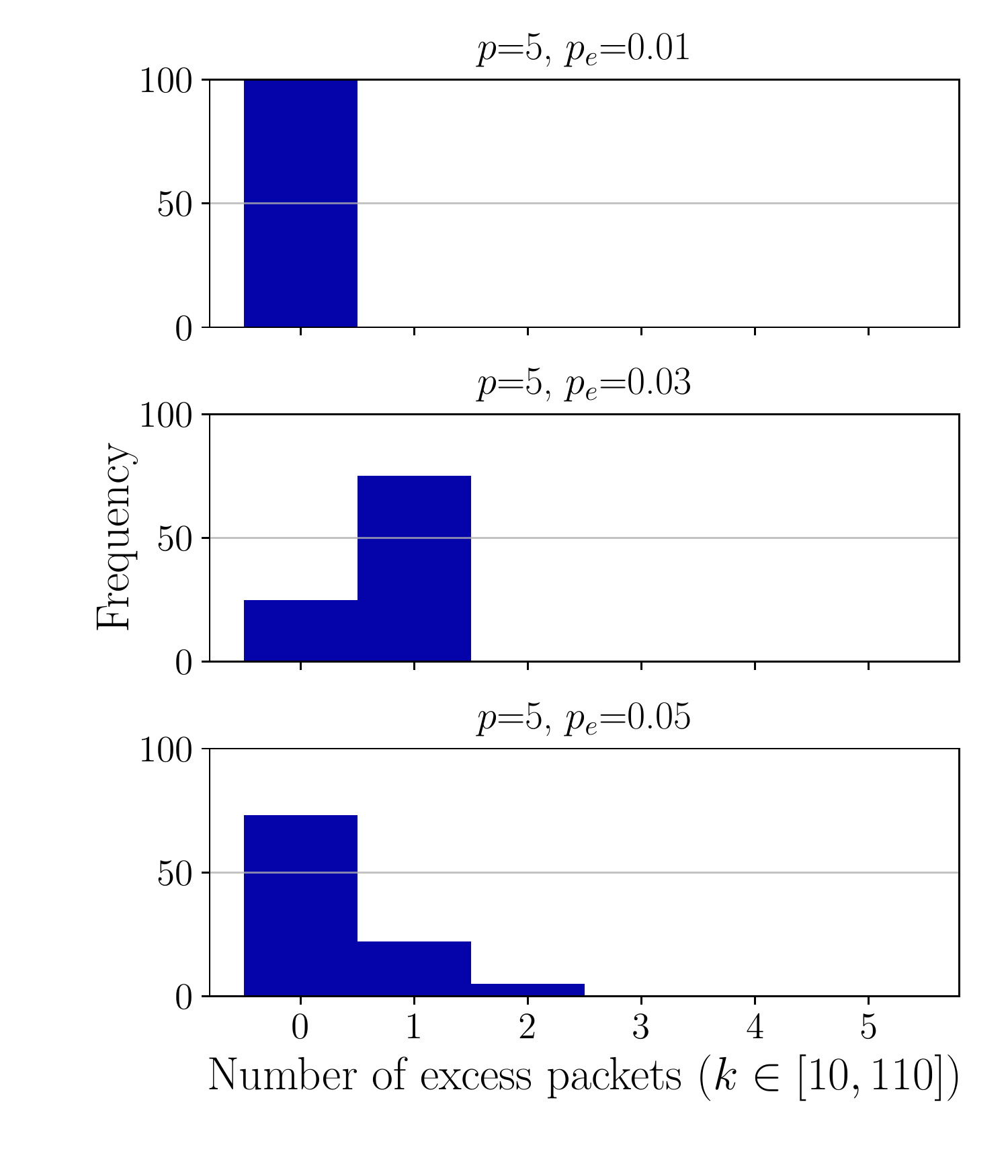}\caption{Required number of excess packets in code partitioning for 5 redundancy
packets ($p=5$) and erasure probability $p_{e}=\{0.01,0.05,0.1\}$,
so that $PLR_{FEC}-PLR_{FEC}^{part}\protect\leq0.001$.\label{fig:excess_packets}}
\end{figure}

~

The analysis of the required excess packets is depicted in Figure~\ref{fig:excess_packets},
which shows three histograms for a fixed number of parity packets
($p=5$) and $p_{e}=\{0.01,0.05,0.1\}$. The three configurations
consider $k\in[10,110]$. In this case, excess packets have alternately
been distributed between both codes until $PLR_{FEC}-PLR_{FEC}^{part}\leq\delta$,
with $\delta=0.001$. As depicted in the picture, in most cases code
partitioning is not penalized by an increase in RI.

\section{Conclusion\label{sec:Conclusion}}

PRRT is an example of a transport protocol using HARQ to provide predictable
reliability and delay, applying an MDS code to generate redundancy
to recover from losses. However, packet encoding and decoding have
high computational complexity, making them unsuitable for energy and
computationally constrained devices. In this paper, we analyze the
complexity of MDS codes and propose code partitioning, an approach
to reduce the said complexity by 50\% with only a slight increase
in the required redundancy information.

\bibliographystyle{ieeetr}
\bibliography{ccnc2021}

\end{document}